\documentclass[aps,prd,twocolumn,showpacs,groupedaddress]{revtex4}
\usepackage{epsfig}
\usepackage{longtable}
\usepackage{amsfonts}

\newcommand{\vmus}{{\rm V}/\mu{\rm s}}
\begin{document}
%\draft
%\preprint{HEP/123-qed}

\title{Fast high--voltage amplifiers for driving electro-optic modulators}
\author{Holger M\"{u}ller}
\email{holgerm@stanford.edu}
\affiliation{Physics Department, Varian Bldg., Room 226, Stanford University, Stanford, CA 94305-4060}
%\homepage{http://www.uni-konstanz.de/quantum-optics}

\date{}

\begin{abstract}
We describe five high-voltage (60 to 550V peak to peak), high-speed (1-300\,ns rise time; 1.3-300\,MHz bandwidth) linear amplifiers for driving capacitive or resistive loads such as electro-optic modulators. The amplifiers use bipolar transistors in various topologies. Two use electron tubes to overcome the speed limitations of high-voltage semiconductors. All amplifiers have been built. Measured performance data is given for each.
\end{abstract}

%\pacs{03.30.+p 04.80.Cc 03.50.De}
%\keywords{}

\maketitle

\section{Introduction}

Electro--optic modulators (EOMs) \cite{VogesPetermann,Hobbs} are used for amplitude, frequency, and phase modulation of laser beams and, for example, as actuators in optical frequency and phase locked-loops. They typically require input voltages in the range of $U_\pi=50-500\,$V for generating an optical phase-shift of $\pi$. Because fast amplifiers capable of driving modulators with these voltages are difficult to make, the overall speed is usually limited by the driver amplifier.

The best (monolithic or hybride) operational amplifiers achieve a slew rate \cite{definitions} of $dU/dt\sim 5000\,\vmus$, but only at low voltages of $\pm 15$\,V. At high voltage, hybride op amps achieve $dU/dt \sim 1000\,\vmus$ at 400\,Vpp (volts peak to peak) output amplitude, i.e., rise-times of $400\,\mu$s and a large signal bandwidth of 0.8\,MHz \cite{TietzeSchenk}. Amplifiers based on discrete MOSFETs \cite{HorowitzHill,Col00} achieve similar performance. For example, amplifiers made by various companies provide up to $800$\,Vpp for sinewaves at frequencies up to 250\,kHz, i.e., $dU/dt=600\,$V/$\mu$s, and twice the voltages in a push-pull configuration, where both electrodes are driven by paraphase signals to double the voltage. Integrated amplifiers for driving CRTs in high-resolution monitors \cite{LM2402} achieve about 60Vpp output voltage and 3ns rise-time for low capacity loads ($<20$\,pF). EOMs can also be driven by RF wide-band amplifiers, preferably in push-pull configuration, but the output voltage of these is generally limited to below 100\,Vpp.

In applications like frequency and phase locked loops for diode-, Titanium:sapphire-, or dye- \cite{HallHaensch} lasers, speed and voltage demands can barely be fulfilled by existing amplifiers; such setups can thus be improved (or simplified) by an improved high-voltage amplifier.

Electrically, an EOM represents a capacitive load of the order of 10-150\,pF. There is a tradeoff between high speed and high output voltage: because the current required to charge the EOM and circuit capacity increases with frequency and voltage, the power dissipation of amplifiers based on a given technology is roughly proportional to the bandwidth times the square of the maximum output voltage; moreover, high voltage semiconductors tend to be slower. In this article, we describe five amplifiers that can drive capacitive loads with different combinations of high speed and output voltage. They use several tricks to achieve high performance at moderate power dissipation. In Sec. \ref{Gael}, we present a push-pull bipolar transistor design which achieves $550$\,Vpp output voltage and $>1.3$\,MHz bandwidth, about twice the one of high-voltage operational amplifiers. Section \ref{LMV11} describes an amplifier whose output voltage and bandwidth of 65\,Vpp and 75\,MHz are comparable to those of high-resolution CRT drivers, but that can drive loads of up to 100\,pF. Section \ref{UHF} describes a nanosecond rise-time push-pull amplifier for 60\,Vpp. At high output voltages, one can achieve improved performance by using electron tubes: Section \ref{PL509} describes an amplifier for a maximum output voltage of $550$\,Vpp and $>5.7$\.MHz bandwidth, more than four times faster than our transistorized amplifier for 550\,V. In  Sec. \ref{QQE0640}, we describe a 140\,Vpp amplifier for $>120$\,MHz bandwidth.

\section{Bipolar transistor designs}\label{bip}

While MOSFETs are intrinsically very fast, their high input and output capacity are significant disadvantages for their use as fast linear amplifiers. For example, an amplifier capable of generating 400Vpp output voltage with 300V/$\mu$s slew rate is described in Ref. \cite{Col00}, intended for the use as a piezo driver. It uses a MTP2P50E (p-channel) and a IRF830 (n-channel) MOSFET, whose output capacities add to 220\,pF. While this is much lower than the capacity of a typical piezo, it is greater than the capacity of a typical EOM (100pF). Thus, when driving an EOM, a large part of the amplifier's potential output current is used to charge the internal capacities. This reduces the potential speed of MOSFET amplifiers. Bipolar transistors, on the other hand, are available that have $<3$\,pF output capacity.

\subsection{550\,Vpp, 300ns differential amplifier}\label{Gael}

This amplifier (Fig. 1) uses a differential amplifier configuration with complementary emitter followers for providing a 550Vpp output voltage with relatively low power dissipation. A positive and a negative signal each provide half the output signal amplitude. With zero input signal, the output voltage across the load is zero. This amplifier can be used for any loads that allow for a differential drive, such as EOMs made by LINOS photonics.

\begin{figure}
\centering \epsfig{file=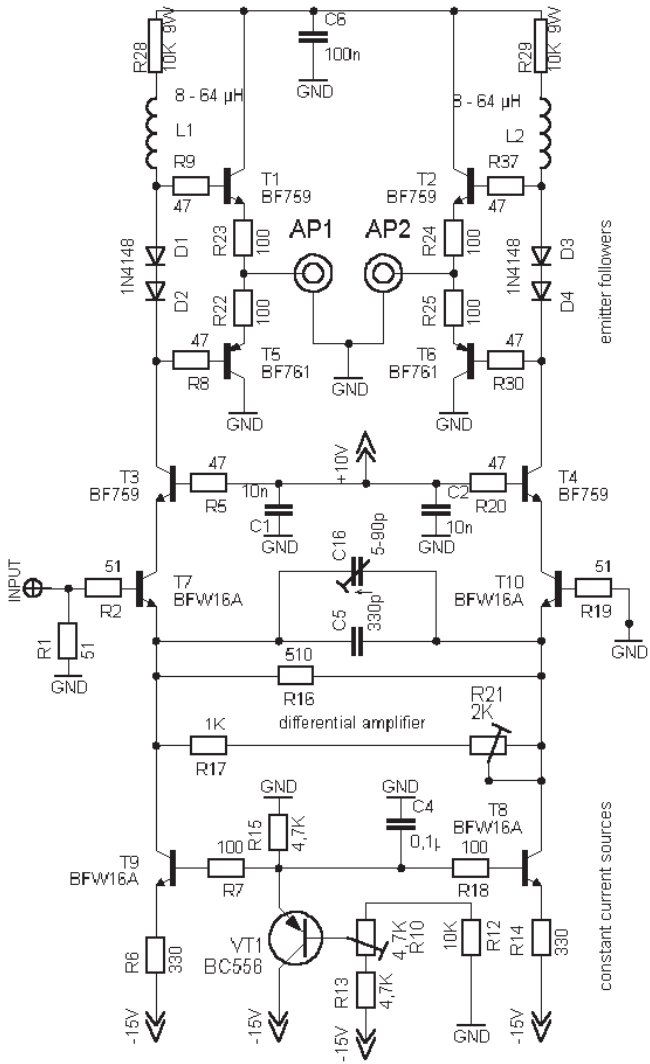, width=0.5\textwidth}
\caption{\label{Gaelschem} Schematic of the 550Vpp, 300ns
amplifier.}
\end{figure}

The main building block is the differential amplifier consisting of T7 and T10. These transistors are connected with T3 and T4, respectively, to form two cascode stages \cite{TietzeSchenk}. T7 and T10 act as current amplifiers. They can be low-voltage types BFW16A that feature a transition frequency $f_t> 1\,$GHz. Voltage amplification is done by T3 and T4, which must be high voltage types. Since they operate in a common-base configuration, their transition frequency $f_t$ can be considerably lower without degrading the overall speed. Moreover, they can be used up to their $U_{cb0}$ voltage rating, which usually is 20 to 50\% higher than $U_{ce0}$, the maximum value for common-emitter configuration. We choose types BF759 and BF761, that feature $U_{cb0}=350$\,V, a maximum total power dissipation $P_{\rm tot}= 10$\,W, $f_T=45$\,MHz and about 3.5\,pF output capacity. (These types seem to be out of production, though they are still available in small quantities. Tested substitutes that even lead to slightly improved speed due to their lower capacities are 2SB1011 (pnp) and 2SC3063 (npn).) T8 and T9 operate as constant current sources for the differential amplifier. The output signals are delivered to the load via two complementary emitter followers, one consisting of T1 and T5, the other of T2 and T6. They are biased for class AB operation by diodes 1N4148 to reduce the crossover distortion.

The gain is set to 100 by $(R_{28}+R_{29})/R$, where $R$ is the resistance of the network consisting of R16, R17, and R21. The bandwidth of the amplifier is determined mainly by the low-pass filter formed by $R_{28}=R_{29}$ and the transistor and wiring capacity at the node at the collectors of T3 or T4, respectively, that amounts to about 15\,pF. This gives a theoretical -3dB--bandwidth $B=1/(2\pi RC)\approx 1.1$\,MHz. The inductors L1 and L2 increase the effective load impedance at high frequencies, which leads to a theoretical increase of the bandwidth by $\sim40\%$ with 1\% overshoot \cite{Valley}. An additional increase of the bandwidth is provided by C5 and C16. The inductors and capacitors are adjusted for optimum square-wave response at a high signal voltage (500Vpp).

R28 and R29 dissipate about 2.25\,W each at zero input signal. However, they must be rated for 10\,W so that the amplifier can be continuously operated with a DC signal of full amplitude. T3 and T4 dissipate at most 2.25W each under all signal conditions; a small heat sink (mounted with minimum stray capacity) is attached to them. Each of T1, T2, T5, and T6 will dissipate 3.75\,W at full output voltage with a 500\,kHz sinewave and 100\,pF load; cooling was provided by attaching the transistors to the metal housing with an insulating layer.

The amplifier was tested with an LM0202 EOM (LINOS AG) as a load, that is specified to have 82\,pF capacity, connected via short (25cm) RG-58 cables (the use of low-capacity types such as RG-63 should be preferred for longer cables). The results are summarized in table
\ref{Gealtech}.

\begin{table}
\caption{\label{Gealtech} Technical data of the 550\,V differential amplifier with a 100\,pF load. With lower capacity loads, speed is up to 50\% larger.  Phase shift is about $45^\circ$ at the -3dB frequency limit.}
\begin{tabular}{lcr} \\ \hline Parameter & Condition &  Value
\\ \hline Gain & & 100 \\
Input impedance  & & 50$\Omega$\footnote{determined by R1} \\ Max. output amplitude  & & 550\,Vpp
\\ Supply & & +300V/80mA \\ & & -15V/30mA \\ & & 10V/20mA \\ Bandwidth &
$U_2\leq 200$\,Vpp & 2.0MHz \\ & $U_2 =500$\,Vpp & 1.3MHz \\ Rise time & $U_2\leq 200$\,Vpp &
150ns
\\ & $U_2 =500$\,Vpp & 300ns \\ Delay & $U_2\leq 200$\,Vpp & 80ns \\ & $U_2 =500$\,Vpp & 180ns
\\ \hline
\end{tabular}
\end{table}

%\begin{table}
%\centering \caption{\label{Gaelresults} Rise-time and delay, measured with a 100\,kHz square-wave
%signal. The bandwidth and the approximate phaseshift between input and output at the -3db
%frequency were measured with sinewaves.}
%\begin{tabular}{ccccc} \hline
%Amplitude & rise time & delay & -3dB frequency & phase shift \\ Vpp & ns & ns & MHz & degrees \\
%\hline 100 & 200 & 80 & 2.0 & 45 \\ 200 & 220 & 85 & 1.9 & 45 \\ 400 & 300 & 150 & 1.2 & 40 \\ 500
%& 400 & 180 & 1.0 & 40 \\ \hline
%\end{tabular}
%\end{table}

\subsection{Single-ended 65\,Vpp, 5ns amplifier with multiple output
stages}\label{LMV11}

EOMs made from materials such as lithium tantalate (LTA) can have $U_\pi <50$\,V. Such voltages can, in principle, be provided by integrated circuits intended for driving cathode-ray tubes in high-resolution monitors \cite{LM2402}. However, while these work well with a low-capacity load of around 10\,pF, they cannot provide the output current of 1.2\,A peak that is required to drive a 100\,pF load with 60\,Vpp at 5\,ns rise time. The amplifier presented in Fig. 2 delivers up to 85\,Vpp into a single ended 100\,pF load at less than 5\,ns rise time.

\begin{figure*}
\centering \epsfig{file=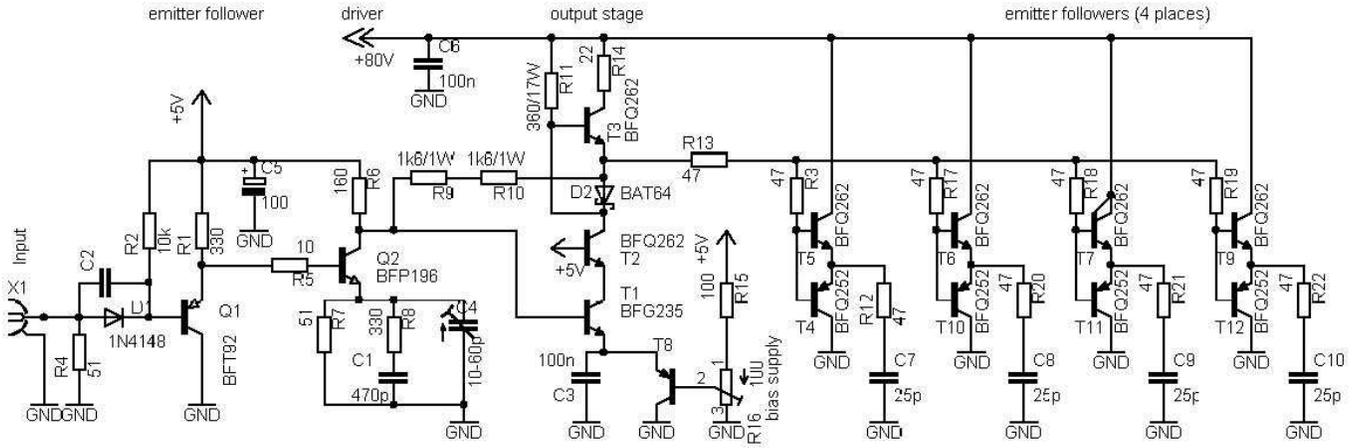, width=\textwidth}
\caption{\label{LMV12} Single-ended 65\,Vpp, 5ns amplifier with
multiple complementary emitter follower output stage. C7-10
represent the load.}
\end{figure*}

Since a single output stage cannot provide the necessary current, we use four complementary emitter output stages in parallel. Each can drive an individual load of 25\,pF. Combining their outputs, a single 100\,pF load can be driven. Without the emitter followers, the amplifier can drive a single 25\,pF load connected directly to the emitter of T3. The output transistors BFQ 262 and BFQ 252, made for the video output stages of high-resolution monitors, have a $V_{cb0}$ of 100\,V (115V for BFQ252A/BFQ262A), $f_T>1$GHz and collector-base capacities $C_{cb}$ of 2.5\,pF (BFQ252) and 2\,pF (BFQ262), respectively. They feature a 5\,W power dissipation. The emitter followers operate without bias in class B, which causes some crossover distortion. When low distortion is important, they can be biased for class AB as the output emitter followers in Fig. 1.

The emitter followers are driven by a voltage amplifier output stage consisting of T1, T2, and T3. T1 and T2 are arranged in a cascode configuration, with the advantages discussed in Sec. \ref{Gael}. The load is connected via the emitter follower T3. Since T3 can only source but not sink current, with a capacitive load (such as the complementary emitter follower output stages) the rise time for negative voltage changes would be much longer than for positive ones. Thus, the Schottky diode D2 has been connected from the emitter of T3 to the collector of T2. For positive voltage changes, D2 is reverse biased, but for negative ones, T2 can sink current from the load through D2. Thereby, negative voltage changes can be as fast as positive without an additional transistor. Alternatively, a complementary emitter follower could be connected to T2's collector, but then the collector-base capacities of three (rather than two) transistors would contribute to the capacity at T2's collector, thus reducing the speed. Circuits of this type are also used for driving cathode-ray tubes in high-resolution monitors.

Adjustable DC bias (nominally 2.3V) is applied by T8 at the emitter of T1. Negative feedback via R9 and R10 sets the gain and gives the stage a low output impedance. A common emitter amplifier Q2 drives the stage. The network at Q2's emitter compensates for the gain loss due to stray capacities parallel to R9 and R10 at high frequencies. The input signal is terminated into 50\,$\Omega$ (R4). A pnp emitter follower Q1 drives the common emitter amplifier Q2; D1 shifts the DC level of the input voltage.

R11 must be mounted with low stray capacity. Since the negative feedback reduces waveform distortions, such as ringing, it may be a wire-wound type. Actually, the inductance of a wire-wound resistor even improves the risetime of this amplifier by series peaking. Under quiescent conditions, T2 dissipates at most 3.5W. At 50MHz, 50Vpp and 25pF load per output, each of the emitter follower output transistors dissipates 2.5W. The amplifier was constructed with all BFQ252 and BFQ262 transistors on a large common heat sink that also served as a ground plane. To minimize the collector capacity of T2, a shield (copper foil) was mounted between two mica insulating layers between heat sink and T2. It was electrically connected to the emitter of T3. The technical data of the amplifier is summarized in Table \ref{LMV12tech}.

\begin{table}
\caption{\label{LMV12tech} Technical data of the 65\,Vpp amplifier.}
\begin{tabular}{lr} \\ \hline Parameter &  Value
\\ \hline Output voltage range & 10-75V\footnote{Can be increased to at
most 10-95V by rising the supply voltage to 100V} \\ Input voltage range & 0-1V
\\ Gain & 60 \\ Input impedance & 50$\Omega$ \\ Load & $4\times 25$pF or $1\times 100$pF \\
Rise time (60\,Vpp) & 5ns \\ Bandwidth & 75MHz \\ slew rate & 3\,kV$/\mu$s \\ Supply\footnote{at 50Vpp, 50MHz and $4\times 25$pF load} & +80V/200mA; +5V/10mA
\\ \hline
\end{tabular}
\end{table}

Without the emitter followers, the output stage of this amplifier uses only npn transistors, which are available with maximum voltages $U_{cb0}>1$kV. For example, two BUX87-1100 transistors using a 700V supply and an output resistance of 12.5k$\Omega$ could generate $>$600Vpp single ended output voltage with $>1$MHz large signal bandwidth.

\subsection{60\,Vpp nanosecond rise-time amplifier with electronic amplitude control}\label{UHF}

The amplifier shown in Fig. 3 is designed for amplifying ECL level pulses, providing 60\,Vpp pulses with nanosecond rise time into a symmetric low capacity load. The output amplitude can be adjusted continuously from zero to 60\,Vpp by a DC voltage.

\begin{figure*}[t]
\centering \epsfig{file=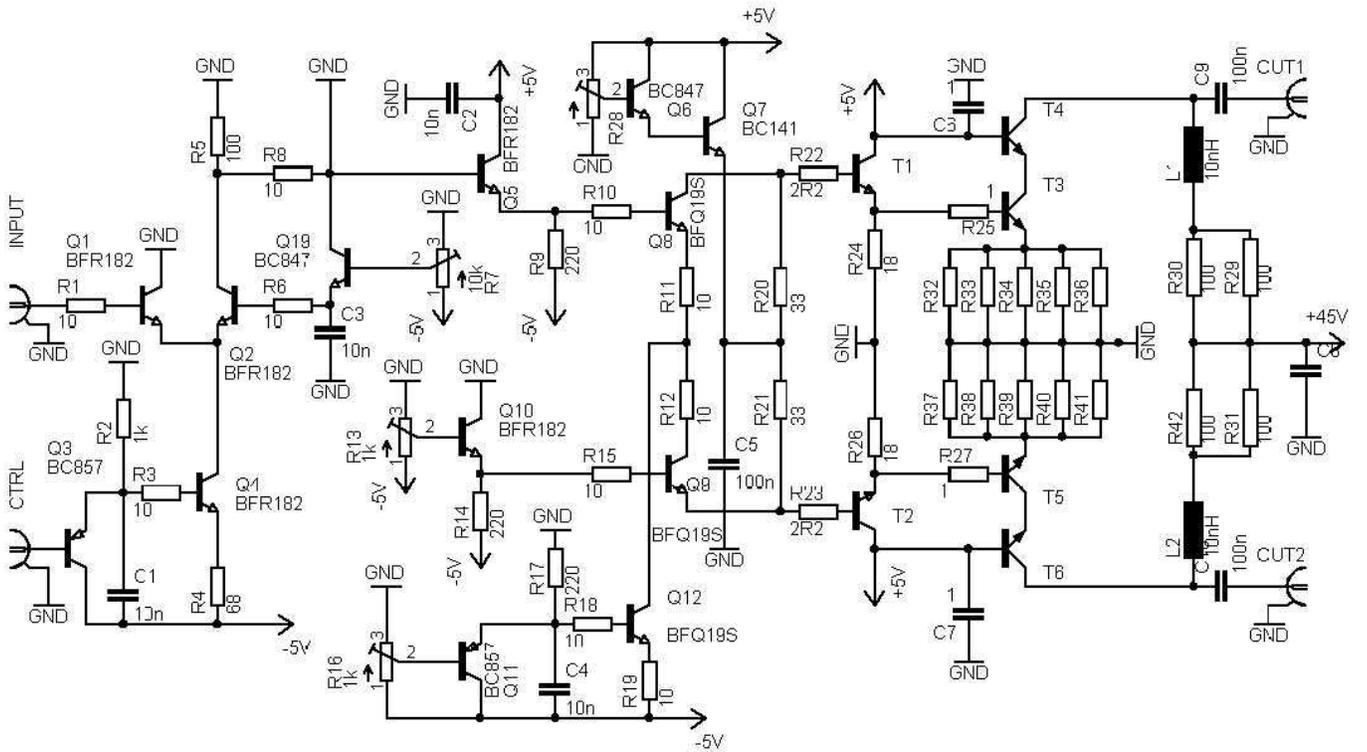, width=\textwidth}
\caption{\label{wuenschmann} Schematic of a 60\,Vpp,
nanosecond-rise-time amplifier.}
\end{figure*}

The input stage is designed for amplitude control. It is a differential amplifier (Q1 and Q2) with an adjustable constant-current source (Q4). The output amplitude is adjustable with a voltage at the base of Q3 between -5\,V and -4\,V. The collector capacity of the BFR182 transistors is only a few tenths of pF, so R5 can have a relatively high value of 100\,$\Omega$, provided that wiring capacities are minimized.

An emitter follower Q5 provides a low impedance signal to the subsequent stage, a differential amplifier consisting of Q8 and Q9. It has threefold voltage gain. Its load resistance must be as low as 33\,$\Omega$, since the more powerful transistors have higher capacities. This stage derives an adjustable supply voltage from Q7. Emitter followers T1 and T2 using the relatively powerful BFG235 transistor drive the power stage. Because of the low impedances here and in the power stage, the layout must be designed for low parasitic inductances. The lead lengths to the bases of T16 and T17 have to be made as short as possible, in order to reach the bandwidth of above 300\,MHz required for the short rise-time. The BFG235s dissipate about 1W through their collector connections.

The output capacity of the power transistors must be as low as 6.6\,pF to achieve the desired speed. They also have to meet a demanding combination of voltage, current, dissipation, and $f_T$ ratings. Only UHF power transistors satisfy these. Many devices, however, come with a relatively narrow band internal impedance matching network and are thus not usable for this application. Transistors BLW33 have been chosen.

To eliminate the Miller effect (which would result in an unrealistically small input impedance), a cascode configuration is the only choice. This has the additional advantage that the ``upper" transistor can be utilized up to $U_{cb0}$, which is crucial because of the relatively low maximum voltages of UHF transistors.

The power stage uses negative feedback by emitter resistors. Each consists of five paralleled resistors (12\,$\Omega$ each) in order to minimize the parasitic inductance to $\ll 1$nH. Also the resistors R29, 30, 31, and 42 must be very low inductance types; suitable are those that come in power transistor packages that can be mounted to a heat sink. L1 and L2 compensate for the gain loss at high frequencies; they consist entirely of resistor and wiring inductances. The power stage dissipates 27\,W. The power resistors and transistors were mounted to heat sink. A fan was used to keep the heat sink below 50$^\circ$C.

The technical data summarized in Tab. \ref{UHFtech} refer to a circuit driven from a high-speed ECL waveform with a rise time $<0.3$\,ns. The load can be connected through two 50\,$\Omega$ coax cables of equal length. They are terminated inside the amplifier, so termination at the load is unnecessary (and would reduce the output voltage to one half).

\begin{table}
\caption{\label{UHFtech} Technical data of the 60\,Vpp nanosecond amplifier.}
\begin{tabular}{lr} \\ \hline Parameter &  Value
\\ \hline Max. output voltage & 0-60Vpp \\ Input voltage & ECL \\
Load & $5$pF \\ Rise time (60\,Vpp) & 1\,ns \\ Short circuit duration & Infinite
\\ Supply & +45V/450mA; +5V; -5V
\\ \hline
\end{tabular}
\end{table}

\section{Electron-tube designs}

Electron tubes are well suited suited for high-voltage linear amplifiers. At a given output capacity, they provide higher power dissipation and current capabilities than transistors (bipolar or MOSFET).

\subsection{Single-ended 550\,Vpp, 60\,ns rise time amplifier}\label{PL509}

The power stage of the amplifier shown in Fig. 4 uses two tubes: V1 to source current to the load and V2 to sink current. The power stage is controlled by the control grid g1 of V2. Under quiescent conditions, this is at about $-23$\,V relative to its cathode. Its corresponding anode current of $I_{a,0}=85$\,mA causes a voltage drop across R2 that causes a similar negative voltage of $-23$\,V at the control grid of V1 with respect to its cathode. If V2 is driven to conduct lower current, there will be less voltage drop across R2, causing V1 to conduct more current. If, on the other hand, V2 is conducting a large current, V1 will be cut off by the voltage drop across R2. Thus, the two tubes act as a push-pull amplifier. The circuit can therefore both source and sink peak currents $I_{a,m}$ of more than 800mA, substantially larger than the quiescent current $I_{a,0}$ \cite{peakcurrentPL509}. This makes a high slew rate $dU/dt$ possible at moderate power dissipation.

\begin{figure*}
\centering \epsfig{file=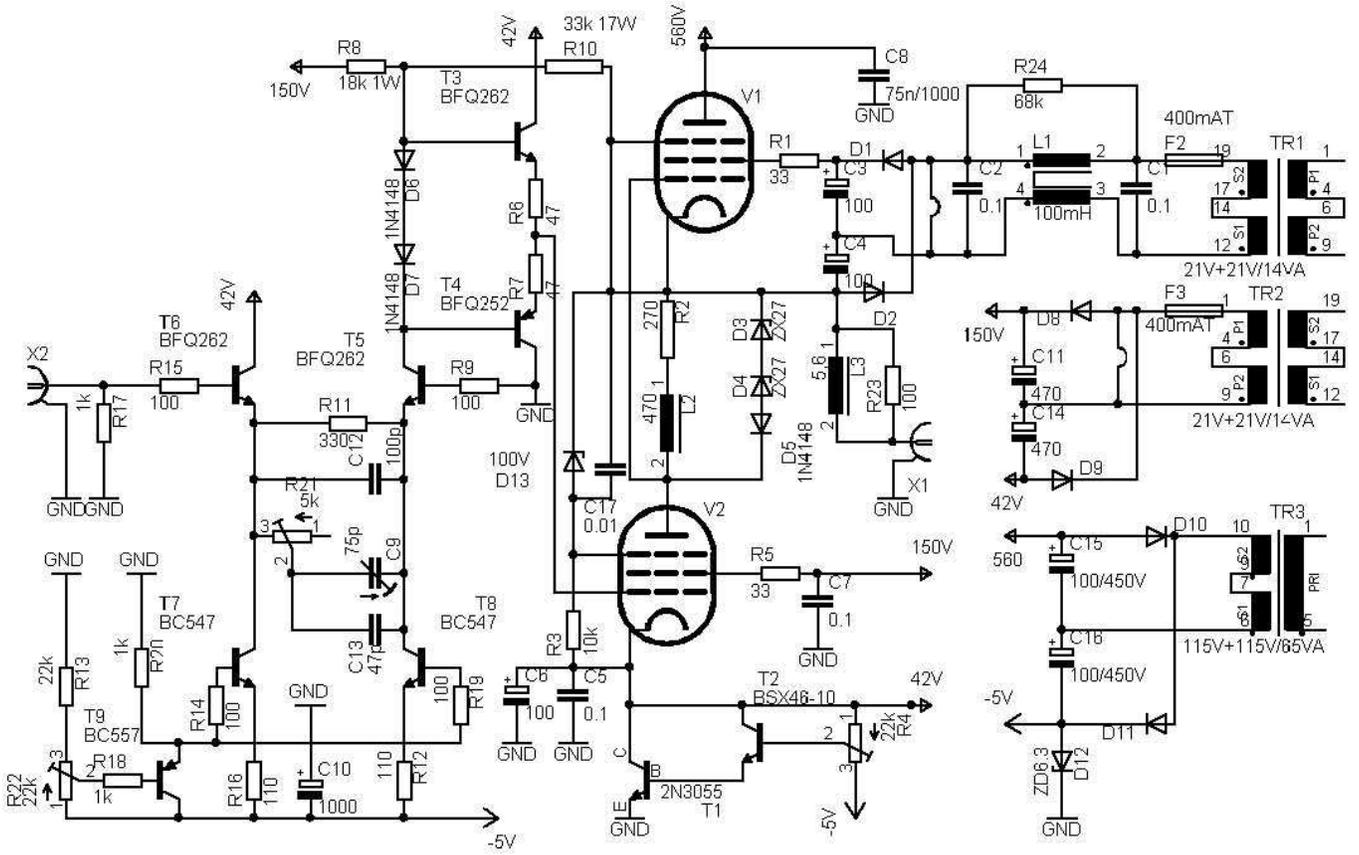, width=\textwidth}
\caption{\label{PL509schem} 550\,Vpp, 100\,ns rise time amplifier
for a single-ended 140\,pF load. The two electron tubes are
PL509.}

\end{figure*}

A differential amplifier (T5 and T6) is used as driver stage. The driver is significantly faster than the power stage. It uses the BFQ252 and BFQ262 described in section \ref{LMV11}. The power stage is driven by a complementary emitter follower (T3 and T4).

Negative feedback over the power stage by R10 is used to set the gain and provide a low output impedance. The output stage without feedback acts like an integrator (providing constant output current to a capacitive load). Since the feedback is over the power stage and the much faster emitter follower, the amplifier does not get instable even with large capacitive loads. A small capacitor in parallel to R10 stabilizes the amplifier. It is effectuated by attaching a thin copper strap on the surface of R10, that is connected to the output side of the resistor. The overall gain of the amplifier is set by $R_{10}/R_{11}$. The high frequency compensation network consisting of C9,C12,C13, and R21 compensates for the distributed feedback capacity parallel to R10. R23 and L3 stabilize the amplifier for capacitive loads.

\paragraph{The g3-trick} The slew rate of the amplifier without load $I_{a,0}/C$ is given by the capacity $C=C_2+2C_{ag1}+C_{\rm stray}$ at the anode of V2, that consists of the output capacity $C_2$ of V2, the anode-to-g1 capacity $C_{ag1}=2.5$\,pF of V1 and V2, and stray capacities $C_{\rm stray}$. The $C_2$ of a PL509 pentode is specified as 17pF with g3 connected to cathode \cite{ValvoPL509,Frank}, so $C\approx 25$\,pF if $C_{\rm stray} \approx 3$\,pF. The output capacity $C_2$ of V2 can be reduced to 9\,pF by the 'g3-trick', i.e., by connecting the g3 of this tube to the cathode via R3 rather than directly, which allows the beam plate g3 to have a floating RF potential. $C_2$ can be further reduced by connecting this g3 to the output (preferably through a 100\,V Zener bypassed with a 10\,nF capacitor to make g3 negative with respect to the anode), thereby practically eliminating the g3-to-anode contribution to the capacity. From the measured rise-time (Tab. \ref{PL509performance}) and $I_{a,0}=85$\,mA, $C\approx 12$\,pF can be calculated, lower by a factor of over 2 compared to the above value, i.e, $C_2$ has been reduced fourfold to about 4\,pF.

\paragraph{The $L_k$ trick} The maximum slew rate of the capacitively loaded amplifier is proportional to the peak anode current $I_{a,m}$, which charges the load and circuit capacities. However, without L2, the slew rate for positive voltage steps at the output would be significantly lower, since the current required to charge the anode capacity of V2 is provided by V1 via R2. This current causes a voltage drop across R2, which makes the grid of V1 negative. This reduces the anode peak current that is available from V1 and leads to a reduced speed of positive transitions. With L2, however, a sudden change in the anode current of V2 cause a positive voltage spike at the grid of V1. In the limit of a very large L2, a small negative change of V2's anode current will drive V1 fully open (in fact, the grid voltage of the upper tube may even get positive). The slew rate for positive edges is thus increased. L2 is bypassed for negative edges by D4 and the 27V Zeners D3 and D4. The Zener is to allow a $\leq 27$V drop across L2 for negative slopes, which is necessary because the arithmetic mean of the voltage across an inductance is always zero.

\paragraph{Power supplies and dissipation} A Voltage-doubler circuit (TR3,D10,D11,C15,C16) provides the positive DC voltages. A negative supply is made by the drop of the power stage's anode current across D12. Regulation of the high--voltage supplies is not used because the feedback across the power stage stabilizes the output voltage in spite of power supply variations.

A floating power supply consisting of TR1 and associated components is used for the screen grid and heater of V1. The ac heater voltage of 40\,V and the screen grid voltage of 100\,V can be generated from a single 40\,V transformer and a voltage-doubler rectifier (D1, D2, C3, and C4). The dual choke L1 prevents the capacity of Tr1's secondary with respect to ground from loading the output by providinbg a high RF impedance. R24 critically damps the series resonance of L1 and the transformer capacity. A similar (non-floating) power supply (TR2 and associated components) produces heater and screen grid voltge for V2. The transistor circuits are also operated from this supply.

T1 and T2 act as shunt voltage regulator to provide an adjustable 42 V bias voltage at the cathode of V2. It uses -5V from the power supply as a reference.

With 400\,Vpp sinusoidal output voltage at 3\,MHz and $C_l=140$\,pF (plus 30\,pF wiring capacities), the power stage draws about 200\,mA average current. Under such conditions, the power supply voltage will decrease to about 500\,V and the power supply drain will be about 100\,W. From these, about 40\,W will be dissipated by each tube, 8\,W by T1, 3\,W by R2, and 5.4\,W each by D3 and D4. The heat is removed by a fan.

\paragraph{Conclusion}
The performance of the amplifier with sine and square-wave signals is summarized in Tab. \ref{PL509performance} and Tab. \ref{PL509tech}. By using other tubes and/or other supply voltages, the circuit can be adapted to a wide range of output voltage, load, and speed requirements. For example, using forced-air cooled tetrodes type 4CX250B, amplifiers with 11\,kV/$\mu$s slew rate are possible for 2000\,Vpp output voltage. On the other hand, downsized versions of the circuit using smaller tubes are also possible. For example, with PL504 tubes (whose g3 is connected internally to the cathode, i.e., the g3-trick unfortunately doesn't work here) at 2000\,V supply voltage, 0.65\.kV/$\mu$s are achieved for 120\,pF loads.

\begin{table}
\centering \caption{\label{PL509performance} Rise and fall time $t_r$, $t_f$, and 3dB-bandwidth $B$ for small ($\leq 200$\,Vpp) and large signals (400Vpp). Phase shift is about $90^\circ$ at the -3dB frequency. With a 250\,pF load, the amplifier will be about 25\% slower than with 140pF.}
\begin{tabular}{cccccc}\\ \hline
$U_2$ & $C_l$ & $t_r$ & $t_f$ & $d$ & $B$ \\
Vpp & pF & ns & ns & ns & MHz
\\ \hline $200$ & 20 & $40$ & $55$ & 40 \\ 400 & 20 & 57 & 60 & 50 & 7.5
\\ $200$ & 140 & $45$ & $90$ & 55 \\
400 & 140 & 60 & 70 & 70 & 5.7 \\ \hline  \end{tabular} \end{table}

\begin{table}
\caption{\label{PL509tech} Technical data of the 550\,Vpp amplifier.}
\begin{tabular}{lr} \\ \hline Parameter &  Value
\\ \hline Output voltage range & 50-600V \\ Input voltage range & $\pm 2.5$V
\\ Gain & 100 \\ Input impedance & 1k$\Omega$ \\ Load & up to 250\,pF \\
Noise (refr'd to input) & 10nV/$\sqrt{{\rm Hz}}$
\\ Hum & 0.5\,Vpp \\ Harmonic distortion\footnote{Measured at 400Vpp with a 140pF load at 1MHz.} & 1.3\%
\\ Max. peak output current & $\pm800$mA \\ \hline
\end{tabular}
\end{table}

\subsection{140V, 120\,MHz driver}\label{QQE0640}

Fig. 5 shows an amplifier capable of driving a low-capacity (10\,pF) electro-optic modulator at $140$\,Vpp with 120\,MHz bandwidth. The unit is built in a small housing that can be directly attached to the EOM, eliminating the cable capacity that would reduce the bandwidth. The dissipated power is removed by a small fan.

\begin{figure}
\centering%\vspace{-1cm}
\epsfig{file=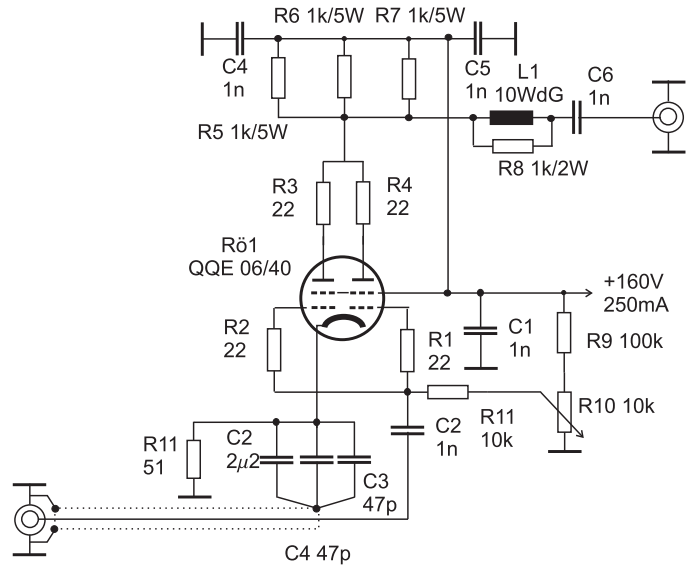,width=0.5\textwidth}
\caption{\label{treiber140MHz} Schematic of the 140\,MHz, 140\,Vpp EOM driver.}
\end{figure}

The QQE06/40 vacuum tube \cite{Frank} is chosen for its low output capacity for a tube of its rated power, and sufficient transconductance of about 20\,mA/V at the operating point (set by R10), giving the amplifier a voltage gain of 16\,dB. The maximum output voltage is about 140\,Vpp with a supply of 180\,V. This is reached with about 30\,dBm, or 20\,Vpp drive level, that can be taken from a commercial semiconductor amplifier.

R3 and R4 prevent parasitic oscillations that might occur at UHF frequencies. It is crucial that C2-4 are connected with minimum lead length from the cathode to the end of the coaxial cable bringing in the signal, rather than to the housing. Otherwise, the resonance of this lead length would decrease the gain strongly above 100\,MHz.

The circuit utilizes a LC compensation network to increase the bandwidth from about $1/(2\pi R_a(C_a+C_{ext}))= 26$MHz (where $C_a=6.4$pF is the output capacity of the QQE06/40, $C_{ext}=12$pF is the sum of the EOM and wiring capacities, and $R_a=330\,\Omega$ is the anode load resistor) to 120\,MHz \cite{Valley}. The inductance of the paralleled wire-wound resistors R5-7 as well as L1 compensate for the gain loss at high frequencies due to the circuit and EOM capacities (both are about $0.9\mu$H). L1 is made of 10 windings of 0.3mm thick copper wire around the 1W metal film resistor R8, which removes a sharp spike in the frequency response that would occur with a lossless inductance here. The LC compensation as shown is optimized for a high bandwidth (140\,MHz, see Fig. 6), but leads to some overshot in the square-wave response and a large phase-shift of 90$^\circ$ at 40\,MHz. For  some applications, like driving an actuator in phase locked loops, a low phase shift is more important. The circuit may also be optimized for low phase shift by setting L1 to zero and increasing the inductance of the anode load resistors to 2$\,\mu$H. This reduces the bandwidth to 60MHz, but at the same time reduces the phase shift to $45^\circ$ at 32\,MHz, corresponding to a delay of $t_d = 4$ns.

\begin{figure}
\centering%\vspace{-1cm}
\epsfig{file=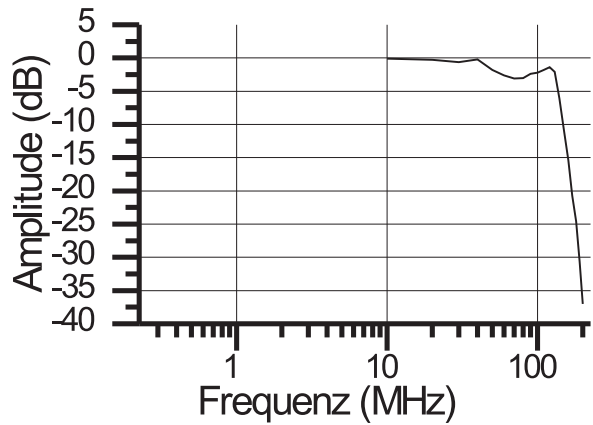,width=0.5\textwidth} \caption{\label{treiber140MHz} \label{freqresponse140MHz} The frequency response of the EOM driver is 3\,dB down at 140\,MHz.}
\end{figure}

The technical data of this amplifier is summarized in table \ref{QQE0640tech}. The load used for this measurement was 10\,pf connected via an SMA connection that adds about 2\,pF.

\begin{table}
\caption{\label{QQE0640tech} Technical data of the 120MHz, 140Vpp amplifier.}
\begin{tabular}{lr} \\ \hline Parameter &  Value
\\ \hline Output voltage & 140Vpp \\ Input voltage & 20Vpp
\\ Voltage gain & 7 \\ Input impedance & 50$\Omega$ \\ Load & 10pF \\
Short circuit duration & Infinite \\ Bandwith & 0.01-120MHz\footnote{may be DC coupled by small
modifications} \\ Phase shift (40MHz) & 90$^\circ$
\\ \hline
\end{tabular}
\end{table}

\section*{Acknowledgments} It is my great pleasure to thank Horst W\"{u}nschmann and Michael Ganslmaier of GW-Elektronik, Munich; Stephan Eggert of the University of Konstanz; Oliver Benson, Claus Palis, and Gael Reinaudi of Humboldt University, Berlin; and Steven Chu, Seokchan Hong, Nate Gemelke, and Edina Sarajlic of Stanford University. This work is sponsored in part by grants from the AFOSR, the NSF, and the MURI. The author acknowledges financial support by the Alexander von Humboldt Foundation.


\begin{references}
\bibitem{definitions} Rise time: time between the 10\% and 90\% points of the rising edge of a square wave; fall time: time between the 10\% and 90\% points of the falling edge of a square wave; delay: time required for the output voltage of an amplifier to reach 50\% of its final value relative to the instant at which a square pulse of negligible rise time is applied at the input. (3dB-) Bandwidth: frequency where the output amplitude of an amplifier driven by sinewaves of constant amplitude is reduced to $1/\sqrt{2}$, or -3dB, of its maximum.
\bibitem{peakcurrentPL509} The available peak current is given by the tube type used and the screen grid voltage. In this circuit, above 800mA are reached with PL509 tubes as specified in the original data sheets \cite{ValvoPL509,Frank}, but some tubes of present day construction provide lower peak current and are thus not suitable.
\end{references}
\end{document}